\documentclass[10pt,letterpaper]{article}
\usepackage{opex3}
\usepackage{color}
\usepackage[colorlinks=true,urlcolor=blue,linkcolor=black,citecolor=black,breaklinks=true,bookmarks=false]{hyperref}

\usepackage{amsmath, amssymb, amsfonts}   
\usepackage{bm}
\usepackage{epstopdf}
\usepackage{here}

\def \widthfig {0.8}
\DeclareMathOperator{\Tr}{Tr}

\begin{document}
	
	\title{Phaseless computational imaging \\ with a radiating metasurface}
	\author{Thomas Fromenteze, Xiaojun Liu, Michael Boyarsky, Jonah Gollub and David R. Smith}
	\address{Center for Metamaterials and Integrated Plasmonics, Department of Electrical and Computer
		Engineering, Duke University, Durham, North Carolina 27708, USA.}
	\email{thomas.fromenteze@duke.edu}

	\begin{abstract}
		Computational imaging modalities support a simplification of the active architectures required in an imaging system and these approaches have been validated across the electromagnetic spectrum. Recent implementations have utilized pseudo-orthogonal radiation patterns to illuminate an object of interest---notably, frequency-diverse metasurfaces have been exploited as fast and low-cost alternative to conventional coherent imaging systems. However, accurately measuring the complex-valued signals in the frequency domain can be burdensome, particularly for sub-centimeter wavelengths. Here, computational imaging is studied under the relaxed constraint of intensity-only measurements. A novel 3D imaging system is conceived based on `phaseless' and compressed measurements, with benefits from recent advances in the field of phase retrieval. In this paper, the methodology associated with this novel principle is described, studied, and experimentally demonstrated in the microwave range.  A comparison of the estimated images from both complex valued and phaseless measurements are presented, verifying the fidelity of phaseless computational imaging.
	\end{abstract}
	
	
	\bibliographystyle{osajnl}
	\bibliography{phaselessbib}

	\section{Introduction}
	
\begin{figure}[h!]
		\centering
		\includegraphics[width=0.7\textwidth]{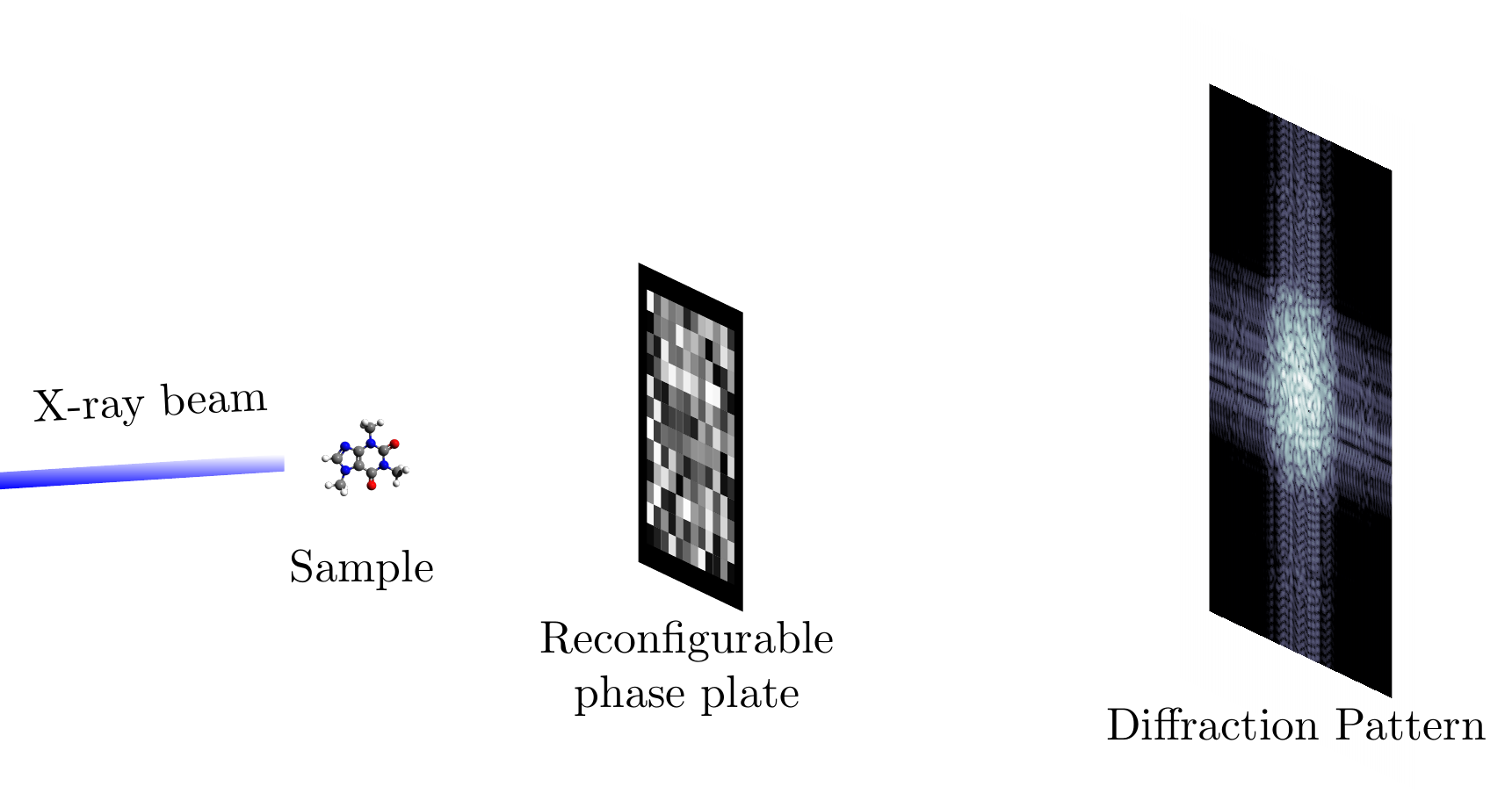}
		\caption{The coded diffraction pattern measured from the illumination of a molecule by an X-ray source. }
		\label{fig:Xray}
	\end{figure}

	Remote sensing exploits the reflection of radiated waves to localize, image, and non-destructively detect objects under study. In imaging applications, measured waveforms are usually back-propagated to the object space using techniques such as Kirchoff migration\cite{zhuge2010modified}, and Stolt's F-K migration~\cite{lopez20003}, giving the location and reflectivity of the target of interest. This process requires the measurement of fast-time varying signals, represented by complex-valued harmonics in the Fourier domain. These phase measurements become challenging in the terahertz, visible, and X-ray ranges, and require the implementation of complex setups, often based on interferometric techniques. Thus, to overcome these hardware limitations, many phase retrieval techniques have been proposed in the last decade to allow for the reconstruction of complex field distributions solely from intensity measurements. They take inspiration from the concept of holograms defined by Gabor~\cite{gabor1948new}, applying alternating projection algorithms introduced by Gerchberg and Saxton~\cite{Gerchberg1972pratical} and Fienup~\cite{fienup1978reconstruction,fienup1982phase}, where at each iteration a complex field source is tailored such that the absolute value of its projection in the target space matches the measured intensity.	In the last decade, a resurgence of attention in the scientific community has focused on the development of efficient phase retrieval algorithms.  Independent and almost simultaneous contributions by Papanicolaou et al.~\cite{chai2010array} and Cand\`es et al.~\cite{candes2015phase} proposed comparable optimization techniques, both based on semi-definite programming of a relaxed problem. The practical implementation proposed by Cand\`es et al. is presented here as the foundation of the approach proposed in this article. This technique focuses on the reconstruction of three-dimensional objects from the measured intensity of coded diffraction patterns (Fig.~\ref{fig:Xray}). A reconfigurable phase plate encodes the fields diffracted from a molecule illuminated by a coherent beam, allowing the reconstruction of a 3D electron density map from multiple intensity measurements of modulated projections. The depicted X-ray imaging setup can be considered under the emerging framework of coherent computational imaging achieved on the physical layer~\cite{hunt2013metamaterial,sun20133d,fromenteze2015waveform}. The aim of this paper is thus to highlight these similarities and demonstrate the reconstruction of near-field 3D images from phaseless measurements taken with a computational system.
	To this end, the mathematical derivations associated with the phaseless measurement of coded diffraction patterns are presented, followed by a concise review of recent phase retrieval techniques. 
	
	The diffracted patterns measured on a plane made of $m$ pixels $\bm y \in \mathbb{R}^m$ are expressed as follows:
	
	\begin{equation}
	\bm y = \lvert \bm A \bm x \lvert ^2
	\label{eq:init}
	\end{equation}  
    
	 where $\bm x \in \mathbb{C}^n$ is an unknown object and ${\bm A \in \mathbb{C}^{m \times n}}$ is a known transfer matrix in which each line $\bm a_i$, \mbox{$i =  1,...,m$} stands for a coded diffraction. The wave diffracted from the object is thus coded by a random mask, giving an illumination pattern $\bm y^{(l)}$ of the form \cite{chen2015solving}:
	
	\begin{equation}
	\bm y^{(l)} = \lvert \bm F \bm D^{(l)} \bm x \lvert^2
	\end{equation} 
	
	where $\bm F \in \mathbb{C}^{n,n}$ is a discrete Fourier transform matrix and $\bm D^{(l)} \in \mathbb{C}^{n,n}$ is a diagonal matrix whose entries are the known random complex transmittance of the mask modulating the diffracted pattern. L random masks are used for the estimation of $\bm x$, leading to $m = nL$ measured points to match the initial formulation of Eq.~\ref{eq:init}. An optimization problem is formulated to find the rank-1 matrix $\bm X = \bm x \bm x^\dagger$. Indeed, for each line $y_k$:
	
	\begin{equation}
		\bm y_i = \lvert \langle \bm a_i, \bm x \rangle \lvert^2 = \Tr(\bm x^\dagger \bm a_i \bm a_i^\dagger \bm x) = \Tr(\bm a_i \bm a_i^\dagger \bm x \bm x^\dagger) = \Tr(\bm a_i \bm a_i^\dagger \bm X)
	\end{equation}
	
	where $\bm a_i \bm a_i^\dagger$ is a rank-1 matrix. $X$ being the outer product of two vectors, this matrix must satisfy:
	
	\begin{equation}
	\bm X \succeq 0,\ \  {\rm rank}(\bm X) = 1,\ \  y_i = \Tr(\bm a_i \bm a_i^\dagger \bm X)\ \text{for}\ i=1,...,m 
	\end{equation}
	
	where $\bm X \succeq 0$ means that $\bm X$ is positive semidefinite and $\dagger$ stands for the conjugate-transpose operator. Because rank minimization is computationally hard, a relaxation was proposed~\cite{candes2015phase} by dropping the rank constraint and replacing it by a trace minimization, accounting for the sum of the singular values of $\bm X$. The semidefinite program PhaseLift is hence defined as:
	
	\begin{equation}
	\begin{aligned}
	& \underset{\bm X}{\text{minimize}}
	& & \mathrm{tr}(\bm X)\\
	& \text{subject to}  
	& & \bm X \succeq 0,\\
	& & &  \Tr(\bm a_k \bm a_k^\dagger X) = y_k, \quad k=1,...,m 
	\end{aligned}
	\end{equation}
	
	The convexity of this formulation makes it solvable with the help of optimization software such as CVX~\cite{grant2008cvx}. Among the numerous applications of this approach, the retrieval from phaseless far-field data of a microwave array has recently been demonstrated in numerical studies~\cite{fuchs2015phase,fuchs2015excitation}. This represents a promising approach for the simplification of radiation pattern characterization and far field imaging systems. However, the dimensionality "lifting" imposed by this algorithm represents the main drawback, squaring the number of unknowns to create the rank-1 matrix $\bm X$.
    
	Alternatively, novel alternating projection algorithms represent an interesting approach for solving large quadratic systems. They demonstrate the efficiency of iterative phase retrieval techniques, such as the block-Kaczmarz method~\cite{wei2015solving}, derived from the algebraic reconstruction technique~\cite{kaczmarz1937angenaherte}, and the Wirtinger flow~\cite{candes2015wirt}. Recently, the truncated Wirtinger flow was proposed by Chen and Cand\`es~\cite{chen2015solving}. It has been demonstrated that these methods always converge to a solution when satisfying support constraints in the spatial domain and appropriate frequency oversampling, in contrast to the most popular methods introduced by Gerchberg, Saxton, and Fienup that can stall in local minima~\cite{candes2015phase}. The truncated Wirtinger flow is the solution adopted in this article for its demonstrated efficiency, although the other mentioned methods remain compatible with quadratic formulations equivalent to Eq.~\ref{eq:init}. This algorithm computes the following equations for each iteration:
	
	\begin{align}
	&\bm x^{(t+1)} = \bm x^{(t)} + \frac{\mu_t}{m} \sum_{i \in S_{t+1} }^{m} \nabla_{l_i}(\bm x^{(t)})\\
	&\nabla_{l_i}(\bm x^{(t)}) =  2\, \frac{\bm y_i - \lvert \bm a_i^* \bm x^{(t)} \lvert^2 }{\bm x^{(t)*}\, \bm a_i}
	\end{align}
	
	For each iteration $t$, the value of $x$ is thus updated by this descent gradient-like algorithm where $\mu_t$ is a step size that can be determined for example by a backtraining line search and $\nabla_{l_i}$ is a descent direction. The algorithm is computed on the adaptive index set $S_{t+1}$ as determined by Chen and Cand\`es~\cite{candes2014solving}, satisfying for any $i \in S_t$:
	
	\begin{align}
	& \bm a_i^* \bm x^{(t)} \approx \lvert\lvert \bm x^{(t)} \lvert\lvert \\
	& \frac{\bm y_i - \lvert \bm a_i^* \bm x^{(t)} \lvert ^2}{\bm a_i^* \bm x^{(t)}} \lesssim \frac{\frac{1}{m} \lvert \bm y_i - \lvert \bm a_i^* \bm x^{(t)} \lvert ^2 \lvert}{\lvert \lvert \bm x^{(t)} \lvert \lvert }
	\end{align}
	
	These constraints improved the efficiency of the initial truncated Wirtinger flow by dropping some gradient components  $\nabla_{l_i}(x)$ bearing too much influence on the search direction. The efficiency of this algorithm is demonstrated in \cite{candes2014solving} considering both ideal and noisy measurements---showing that the computational effort required for solving a random quadratic system is on the same order of magnitude of that of a linear system of the same size.
	
	\section{Phase retrieval adapted to near field imaging using a radiating metasurface}
	
	We propose the adaptation of the phase retrieval framework to a microwave computational imaging system by replacing the coding phase plate presented in the depicted X-ray system (Fig.~\ref{fig:Xray}) by a radiating metasurface and implementing the truncated Wirtinger flow instead of the PhaseLift presented earlier. The codes made openly available by their authors, Chen and Cand\`es, are adapted to this study. Several studies have demonstrated the possible simplification of imaging hardware exploiting the recent development of spatially resolving antennas~\cite{marks2016spatially}. In this context, systems radiating structured illumination patterns have been proposed, exploiting the inherent structure in the imaged targets to limit the hardware complexity~\cite{hunt2013metamaterial,sun20133d,fromenteze2016single}. To this end, frequency diverse~\cite{fromenteze2015waveform,fromenteze2015computational} and dynamically reconfigurable radiating apertures~\cite{sleasman2015dynamic} were studied, demonstrating the linear dependency between the target space and the measured signals through tailored sensing matrices. In this paper, the principle of computational imaging is thus adapted to the phase retrieval problem using a frequency diverse radiating structure, taking benefit of the hardware simplification allowed by both approaches to propose a new paradigm for imaging systems. This demonstration is proposed in the microwave range as a proof of concept to easily compare the reconstructions from complex valued and intensity measurements, paving the way for millimeter wave, terahertz, and photonic applications.
    
\begin{figure}[h!]
		\centering
		\includegraphics[width=0.6\textwidth]{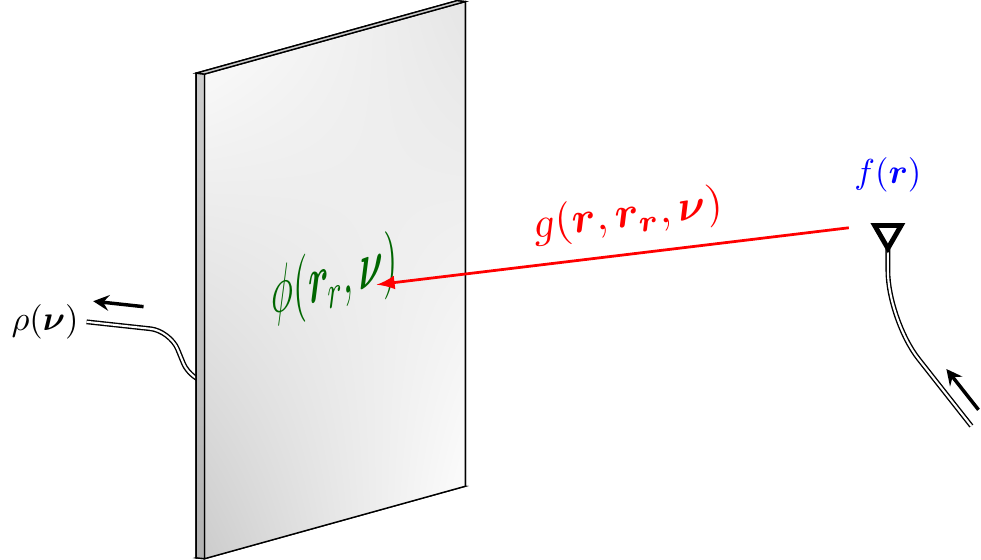}
		\caption{Computational imaging system used for the experimental demonstration. A metasurface radiating frequency diverse patterns is applied to the localization of field sources from the intensity measurement of a compressed frequency domain waveform.}
		\label{fig:Cavity}
	\end{figure}

	Here, we study the mathematical formalism and develop the conditions in which intensity measurement of compressed waveforms can be applied to retrieve the positions of targets in the near field. The experimental setup and the associated parameters are defined in Fig.~(\ref{fig:Cavity}).
	
	In this setup, a frequency diverse structure similar to those introduced in~\cite{hunt2013metamaterial,fromenteze2015computational,lipworth2015comprehensive,yurduseven2016printed}
	is considered. The large modal diversity excited by these metasurfaces allows for the radiation of structured field patterns which vary with the driving frequency. The quality factor of the metasurface is optimized to avoid modal degeneration, allowing for the radiation of a large number of pseudo-orthogonal spatial modes sensing the target space. The expression of the measured frequency signal $\rho(\nu)$ on the radiating device's output port can be expressed as:
	
	\begin{equation}
	\rho(\nu) = \int_{r_r} \int_{r} \phi(r_r, \nu)\ g(r_r,r,\nu)\ f(r)\ d^3 r\ d^2 r_r
	\label{eq:formulation}
	\end{equation}
	
	where $\phi(r_r, \nu)$ stands for the near field distribution of the metasurface measured at the aperture locations $r_r$ for each frequency $\nu$, $g(r_r,r,\nu)$ represents the Green function modeling the propagation of field from the object space $r$ and the metasurface's aperture, and $f(r)$ corresponds to a field source that is localized with this computational system. This problem can be expressed using a matrix formalism by discretizing equation Eq.~\ref{eq:formulation}; we represented the resulting vectors and matrices in bold notation as $\bm r = [r_i]_{1 \leq i \leq n}$, $\bm r_r = [r_{r_j}]_{1 \leq j \leq n_r}$, and $\bm \nu = [\nu_k]_{1 \leq k \leq m}$. A sensing matrix $\bm H \in \mathbb{C}^{m \times n}$ is defined by the product of the Green function matrix $\bm G_{n,n_r}(\nu_k)$ and the cavity near-field response written in the vector form $\bm \phi_{n_r}(\nu_k)$ for each frequency $\nu_k$:
	
	\begin{equation}
	\bm H_{n}(\nu_k) = \bm G_{n,n_r}(\nu_k)\ \bm \phi_{n_r}(\nu_k)
	\label{eq:Hmat}
	\end{equation} 
	
	The sensing matrix allows for a representation of the linear dependency between the measured frequency signal $\bm \rho \in \mathbb{C}^m$ and the object $\bm f \in \mathbb{C}^n$, leading to the following formulation:
	
	\begin{equation}
	\bm \rho = \bm H \bm f
	\label{eq:computational}
	\end{equation}
	
	Previous works demonstrated methods of reconstructing the image vector $\bm f$ under certain invertibility conditions of sensing matrix $\bm H$, corresponding to a sufficient number of radiated orthogonal patterns interrogating the target space~\cite{hunt2013metamaterial}. This work extends the frame of computational imaging by considering the measurement of the intensity of the compressed waveform $\bm \rho$ , described by the following quadratic equation:
	
	\begin{equation}
	\lvert \bm \rho \lvert ^2 = \lvert \bm H \bm f \lvert ^2 
	\label{eq:SigSqrd}
	\end{equation}
	
	Similarly to the coded diffraction problem, an object is reconstructed from phaseless data knowing the complex transfer function of the coding system. However, the frequency diversity exploited by this approach coupled to the associated derivation makes it distinguishable for its compatibility to near-field imaging, without using any actively reconfigurable radiating components. A simulation of the system depicted in Fig.~\ref{fig:Cavity} is studied to identify the number of independent modes required to ensure the reconstruction of the object under study.
	
\begin{figure}[h]
		\centering
		\includegraphics[width=0.7\textwidth]{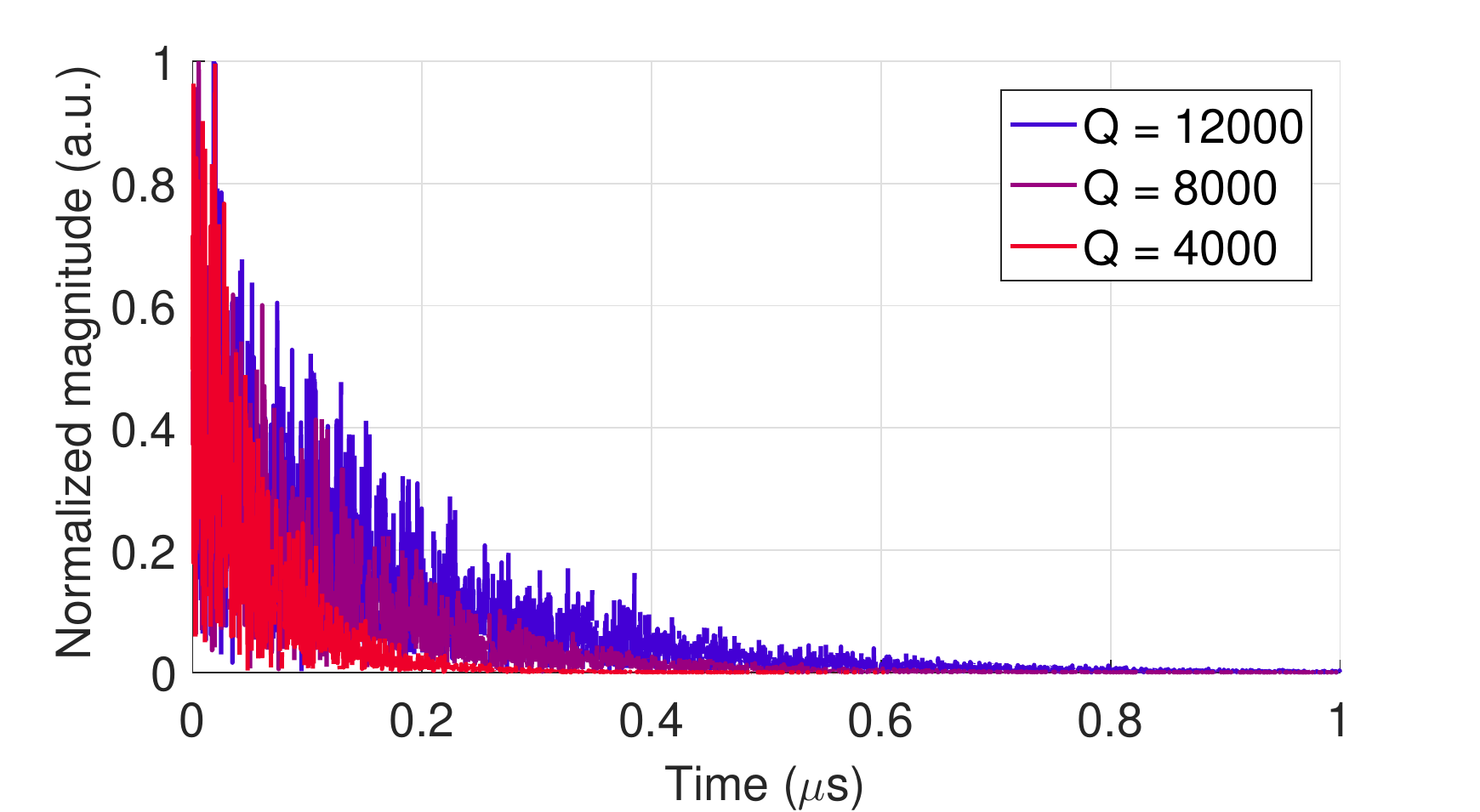}
		\caption{Impact of quality factor $Q$ on the damping time of the metasurface random responses.}
		\label{fig:Q}
	\end{figure}

	In the numerical model of the frequency dependent and randomized radiation pattern of the metasurface, the radiating plane is set at $y=0$, where $y$ is the propagation axis. In accordance with ~(\ref{eq:formulation}), the radiation $f(\bm r)$ of a field source (set in the Fresnel zone of the radiating metasurface), is propagated to the aperture plane $\bm r_r$, compressed by its near-field responses $\phi(\bm r_r, \bm \nu)$ into a unique measurement at the port where the phaseless measurement is performed. Careful consideration of the dispersive nature of the metasurface $\phi(\bm r_r, \bm \nu)$ is required to study the convergence of the computational phase retrieval problem. In the numerical simulations, the impulse response is defined in the radiating plane by a Gaussian random distribution $\mathcal{R}_d(\bm r_r,\bm t)$ with mean value zero, variance one, and discretized over steps of $c/2 \nu_{max}$ in space and $1/B$ in time. The quality factor $Q$ of the metasurface is included in this model to consider the cavity's intrinsic losses and the coupling with all of the irises, leading to a modal degeneration~\cite{marks2016spatially}. The random field distribution is thus multiplied by a decaying envelope or damping time $\tau = Q/\pi\nu_0$, $\nu_0$ being the central frequency of the studied bandwidth (Fig.~\ref{fig:Q})~\cite{marks2016spatially,fromenteze2016single}.
	
	Then, computing the Fourier transform from the time domain to the discrete set of frequency components $\nu$, the full model of the metasurface spatial and frequency response is:
	
	\begin{equation}
	\phi(\bm r_r, \bm \nu) = \mathfrak{F} \big[\mathcal{R}_d(\bm r_r,\bm t)\ \exp(-\bm  t\ \pi \nu_0 /Q) \big]
	\label{eq:meta}
	\end{equation}
	
	In coherent computational imaging system, where complex-valued signals are measured, dispersive antennas presenting long lasting responses are designed to limit the correlation between each radiation pattern in the frequency domain, improving the conditioning of the sensing matrix $\bm H$. In this way, an estimation of the target signature $\hat{\bm f}$ can be computed following $\hat{\bm f} = \bm H^+ \bm \rho$, where the superscript $+$ stands for the Moore-Penrose pseudo-inverse operator~\cite{fromenteze2015unification}. An ideal metasurface would presented perfectly orthogonal radiation patterns ensuring that $\bm H^+ \bm H = \bm I$. In practice, metasurfaces are designed to have low correlation among patterns exploiting the frequency diversity, leading to a non-ideal inversion of the sensing matrix and the apparition of parasitic sidelobes. If we are limited to the measurement of intensity $\lvert \rho(\bm \nu) \lvert^2$, such a direct approach can not be applied. But, alternating projection algorithms may be adapted to this problem to determine of the impact of a radiating metasurface's characteristics. 
	
	The spatial domain sampling is determined by the dimensions of the metasurface, modeled by an array of frequency dispersive dipoles with responses $\phi(\bm r_r, \bm \nu)$. According to the Rayleigh resolution limit, the transverse spatial sampling for a radiating array of dimensions $D_x$ and $D_z$ at a distance $R$ is $d_x = \lambda_c R/D_x$ and $d_z = \lambda_c R/D_z$. The range sampling for a wideband system is given by the width of the radiated pulses as $d_y = c/(2B)$, where $B = \nu_{max}-\nu_{min}$ is the operating bandwidth.
	
	Assuming the measurement of the intensity of a frequency signal $\bm \rho$ described by~(\ref{eq:SigSqrd}), a numerical study is carried out to estimate the criteria required for an accurate reconstruction of a spatial field distribution $\hat{\bm f}_I$, where the subscript $_I$ denotes a reconstruction from an intensity measurement. The performance of such a computational system can be validated against a reconstruction of complex-valued measurements, $\hat{\bm f}$, with a relative error computed for each simulation as:
	
	\begin{equation}
	\epsilon = \min_{\theta \in [0,2\pi]} \frac{\lvert\lvert \hat{\bm f}_I - \hat{\bm f} e^{j\theta} \lvert\lvert}{\lvert\lvert \hat{\bm f}  \lvert\lvert}
	\end{equation}
	
Because the reconstruction from phaseless techniques is unable to estimate an absolute phase, a rotation of the samples by $\theta = \langle \hat{\bm f_I},\hat{\bm  f} \rangle$ is performed to align the estimations in the complex plane before the subtraction. According to~\cite{wei2015solving,candes2015wirt}, successful estimations are considered when $\epsilon \leqslant 10^{-5}$. The field source domain is defined by a number of voxels of positions $\bm r$. The ratio between the number of measured modes $m$ and the number of reconstructed voxels $n$, determines the size of the sensing matrix $H(\bm r, \bm  \nu)$. The ratio $m/n$ is considered in this study for different values of $Q$ to determine the relation between the number of  intensity measurements and the size of the reconstructed domain. The simulation is performed on a frequency band defined between $\nu_{min} = 17.5$ GHz and $\nu_{max} = 26.5$ GHz (K-band), with a frequency sampling $d\nu = (\nu_{max} - \nu_{min})/m$.
	
\begin{figure}[h!]
		\centering
		\includegraphics[width=0.65\textwidth]{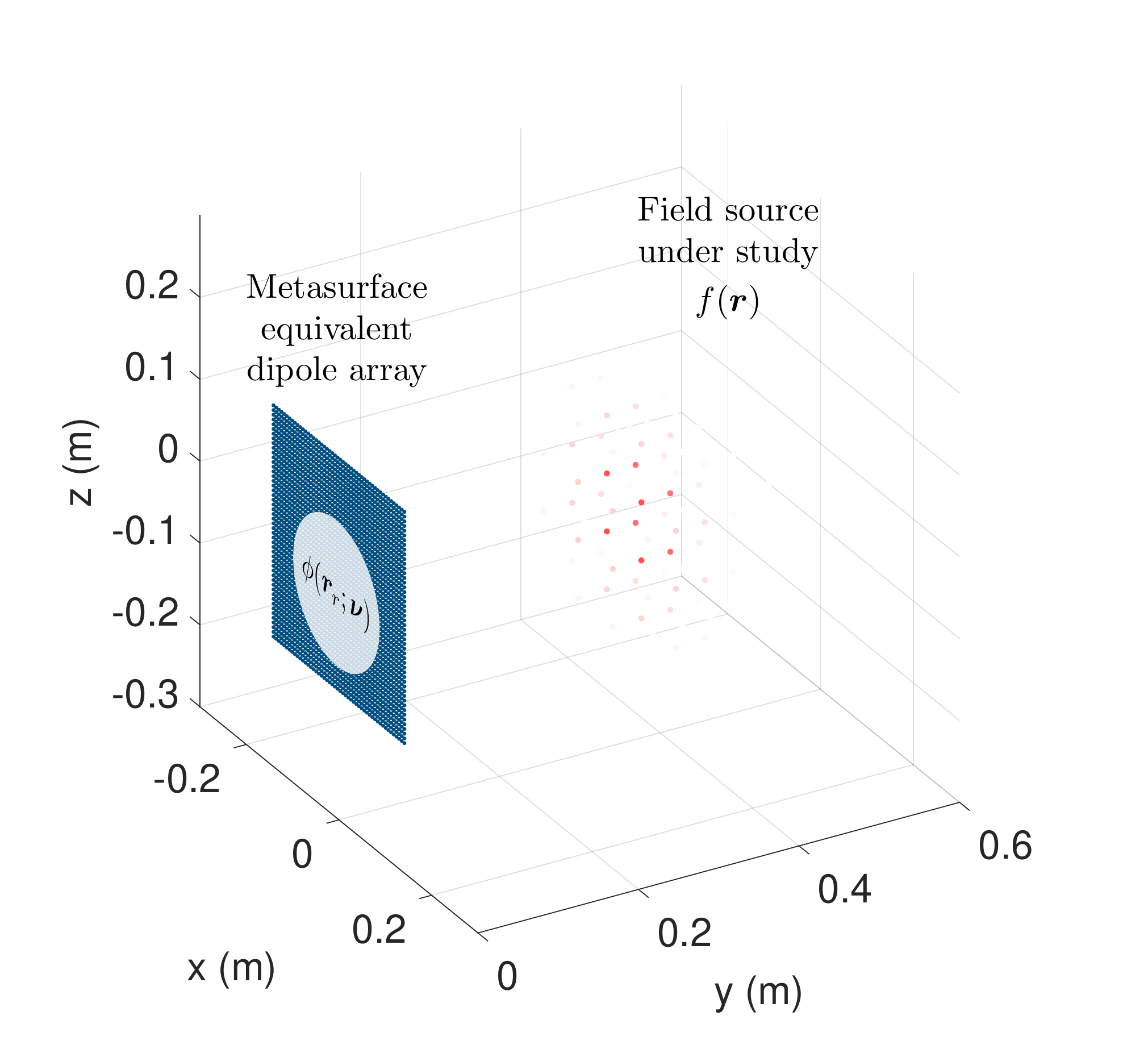}
		\caption{Numerical simulation of the phaseless computational system applied to the localization of a field source. The radiated field is propagated to the receiving structure plane, coded by the response of this metasurface and compressed into a unique frequency domain signal.}
		\label{fig:simulationProbe}
	\end{figure}

	The metasurface delay spread $\tau$ and the equivalent quality factor $Q$ are defined according to the frequency sampling as:
	
	\begin{align}
	\tau &= \frac{\alpha_t}{d\nu} \label{eq:tau}\\
	Q &= \alpha_t \pi \frac{\nu_0}{d\nu} \label{eq:Q}
	\end{align}
	
	where $\alpha_t$ is a frequency sampling parameter set according to the metasurface's damping factor $\tau$. According to~\cite{marks2016spatially}, the optimal sampling is $\alpha_t = 1/ \pi \approx 0.32$ considering that at most $\pi \tau B$ orthogonal channels can coexist due to modal degeneration. The study is presented for $\alpha_t \in [0.1,2]$ to demonstrate the impact of frequency averaging on the phase retrieval algorithm. A domain of $n = 6^3 = 216$ voxels is considered in which a complex random field $\bm f$ must be retrieved. An array of $20 \times 20$ radiating dipoles spatially spaced at $\lambda_{min}/2$ in both transverse directions is considered to simulate a metasurface whose frequency response is defined by~(\ref{eq:meta})~(Fig.~\ref{fig:simulationProbe}).
	
\begin{figure}[h!]
		\centering
		\includegraphics[width=0.65\textwidth]{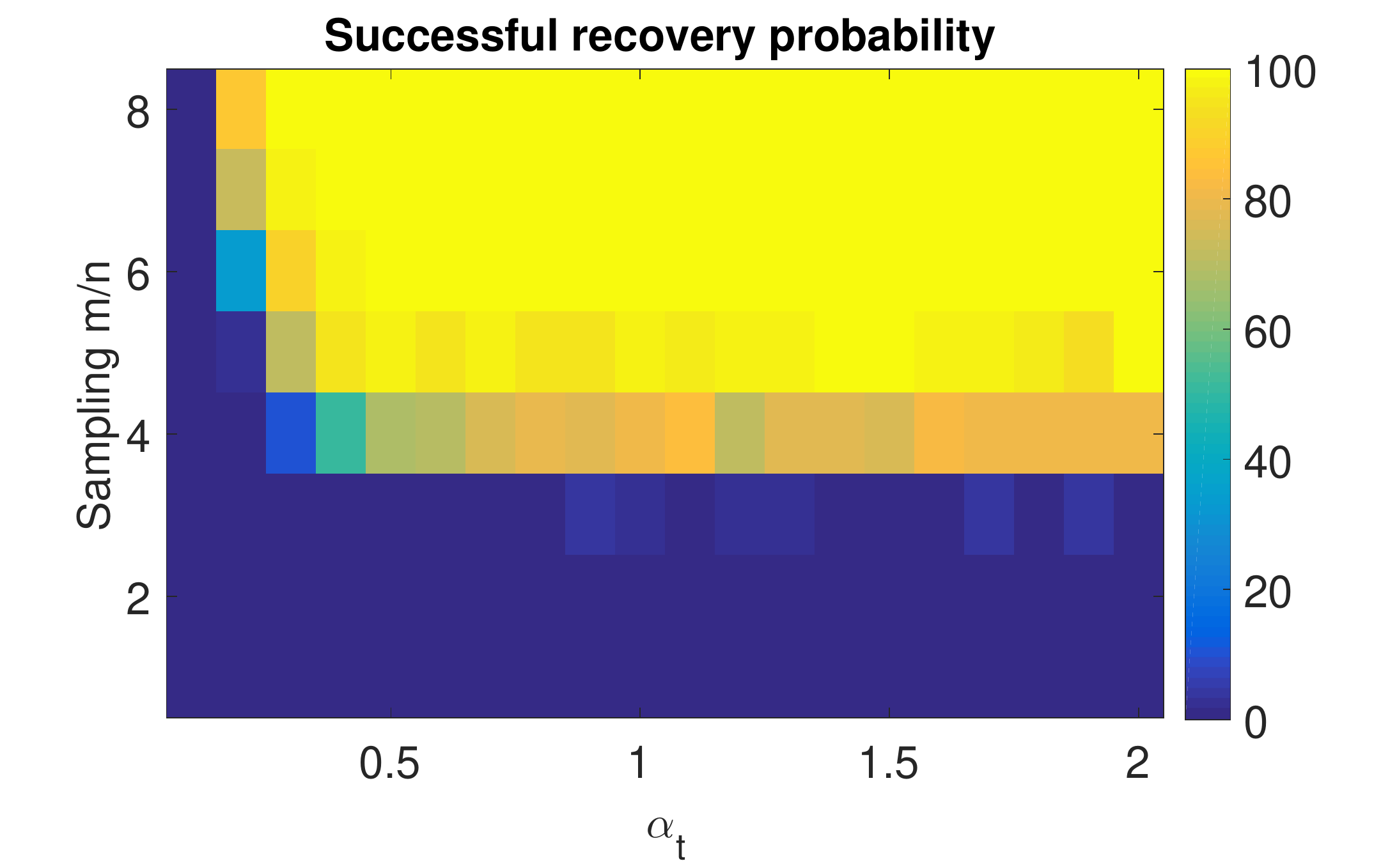}
		\caption{Empirical probability of successful recovery. 100 trials are simulated for each pair of parameters $m/n$ and $\alpha_t$.}
		\label{fig:QvsOversamp}
	\end{figure}

\begin{figure}[H]
		\centering
		\includegraphics[width=0.62\textwidth]{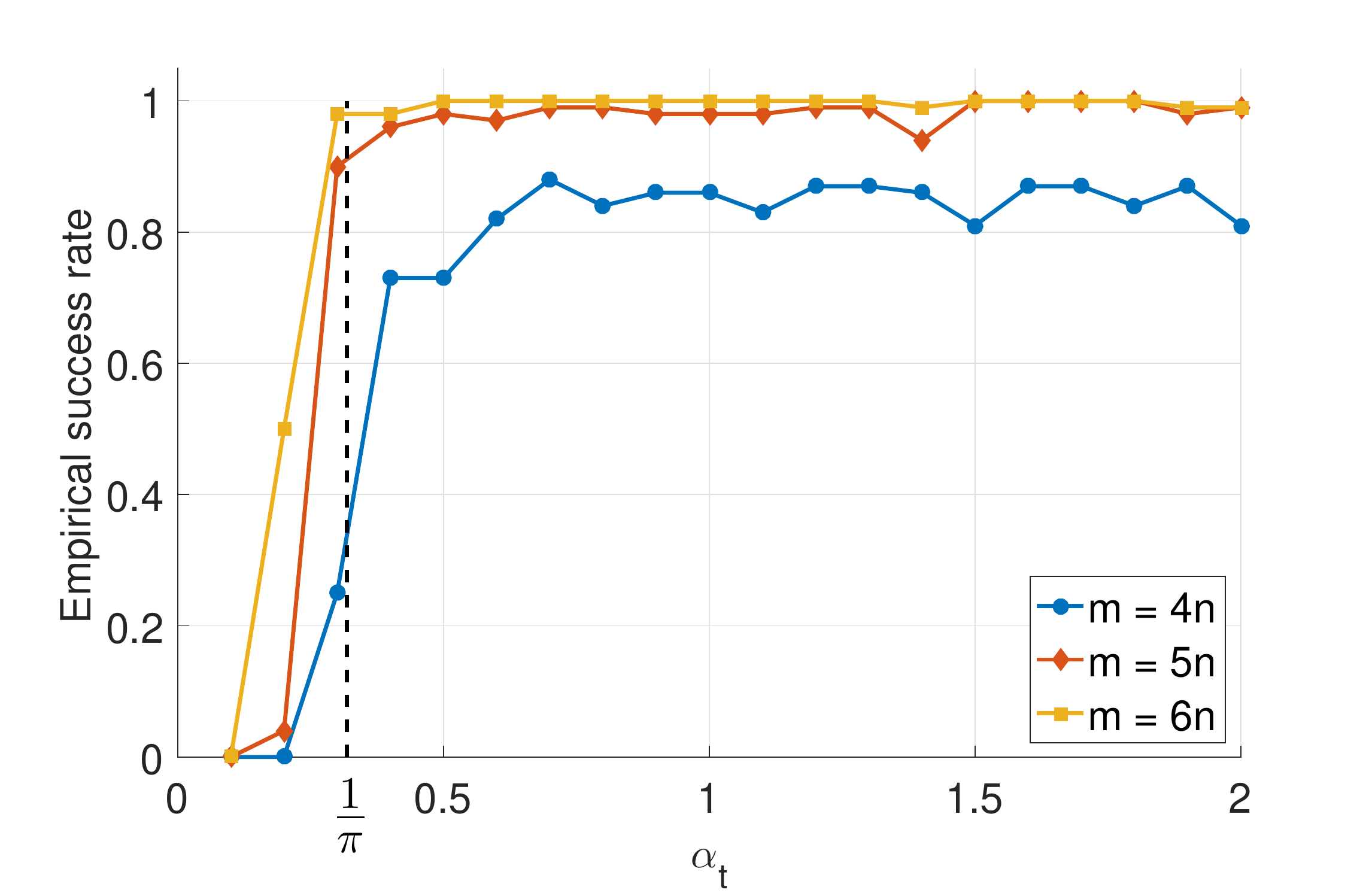}
		\caption{Empirical success rate according to the sampling $m/n$. $\alpha_t$ must be larger than $1/\pi$ to ensure that there is at least as many measured modes $m$ as orthogonal channels available in the sensing matrix $H$ to reconstruct $n$ voxels.}
		\label{fig:convPlot}
	\end{figure}

    100 trials with random metasurface responses and field distributions are computed for each couple of parameters to estimate an empirical set ensuring the accurate estimation of $\hat{\bm f_I}$~(Fig.~\ref{fig:QvsOversamp}). According to this numerical analysis, the probability of successful recovery tends to reach its maximum when $m/n \geqslant 5$ and $\alpha_t \geqslant \frac{1}{\pi}$~(Fig.~\ref{fig:convPlot}). 
		
	As predicted in~\cite{candes2014solving}, a sampling of a least $m = 5n$ ensures a good agreement between the estimations of the field source $\bm f$ with and without the phase information. Furthermore, the relation between the quality factor of the designed metasurface and the frequency vector $\bm \nu$ must satisfy:
	
	\begin{align}
	\alpha_t = \frac{Q d\nu}{\pi \nu_0} &\geqslant \frac{1}{\pi}
	\end{align}
	
	Considering the case of a minimal sampling $m = 5n$ and the definition of the frequency sampling $d\nu = B / m$, the maximum number of voxels imaged with this technique takes the following form:
	
	\begin{align}
	n \leqslant \frac{QB}{5 \nu_0}
	\end{align}
	
	Having identified the critical designing and sampling parameters allowing for accurate estimation of the compressed, phaseless measurement, the impact of additive noise is studied in the next section to evaluate the performance of the algorithm in the context of near field computational imaging. The model given by Eq.~\ref{eq:computational} is modified to account for a Gaussian additive noise $\bm \eta$:
	
	\begin{align}
	\lvert \bm \rho \lvert^2 &= \lvert H \bm f + \bm \eta \lvert^2
	\end{align}
	
\begin{figure}[h!]
		\centering
		\includegraphics[width=0.9\textwidth]{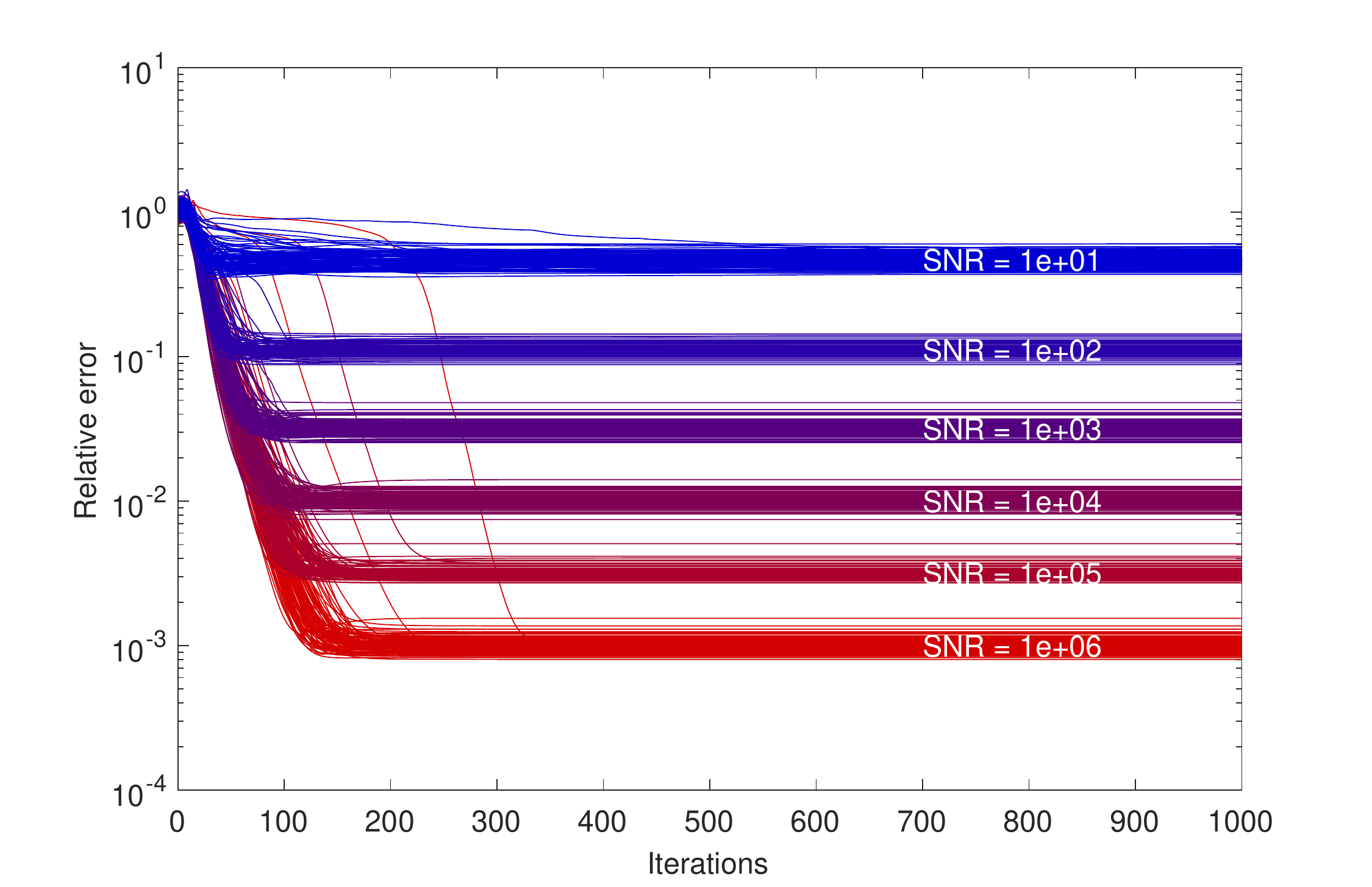}
		\caption{Convergence of the phase retrieval algorithm according to the SNR for $m/n = 6$ and $\alpha_t = 2$. 100 trials are simulated for each SNR value.}
		\label{fig:ConvergenceSNR}
	\end{figure}

	With a sampling $m = 6n$ and for different values of signal-to-noise ratio (SNR), 100 trials are computed with random field distributions $\bm f$ and metasurface responses characterized by $\alpha_t = 2$ to study the convergence of the truncated Wirtinger flow adapted to this near field computational imaging problem (Fig.~\ref{fig:ConvergenceSNR}). 
    
\begin{figure}[h!]
		\centering
		\includegraphics[width=0.9\textwidth]{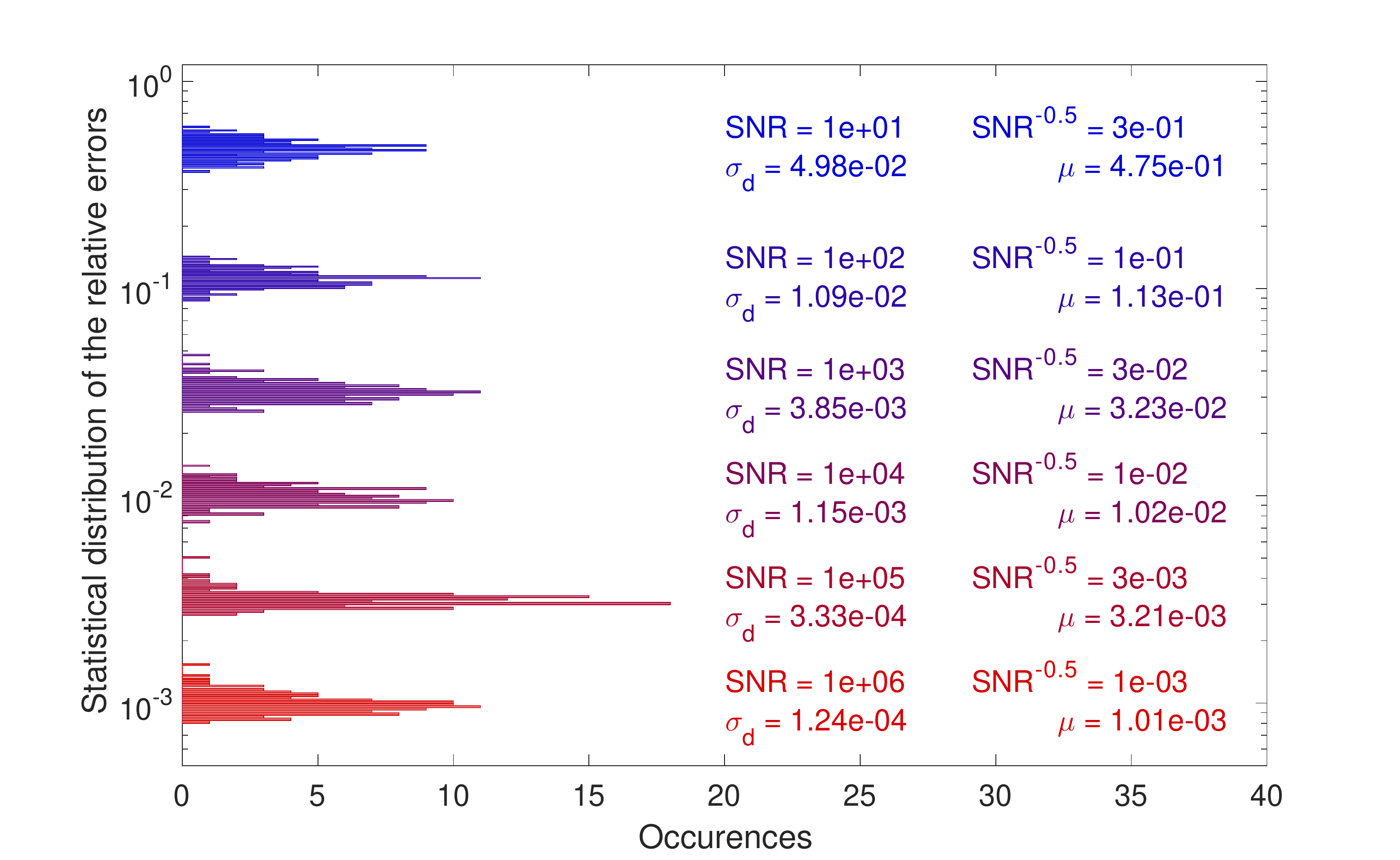}
		\caption{Statistical distribution of the relative errors according to the SNR. In each case, the average $\mu$ of the relative error converges to the normalized noise floor $1/\sqrt{\text{SNR}}$. The standard deviation of each distribution is given by $\sigma_d$.}
		\label{fig:ConvStat}
	\end{figure}

    The value of $\alpha_t$ is deliberately chosen to be large in order to speed up the numerical convergence. According to the sampling $m/n$, the algorithm converges on average in less than 200 iterations. For each value of SNR, the average $\mu$ and standard deviation $\sigma_d$ is computed~(Fig.~\ref{fig:ConvStat}). According to the SNR value, the phase retrieval algorithm demonstrates its efficiency by leading to an average relative error $\mu$ of $1/\sqrt{\text{SNR}}$.
	
\begin{figure}[H]
		\centering
		\includegraphics[width=1\textwidth]{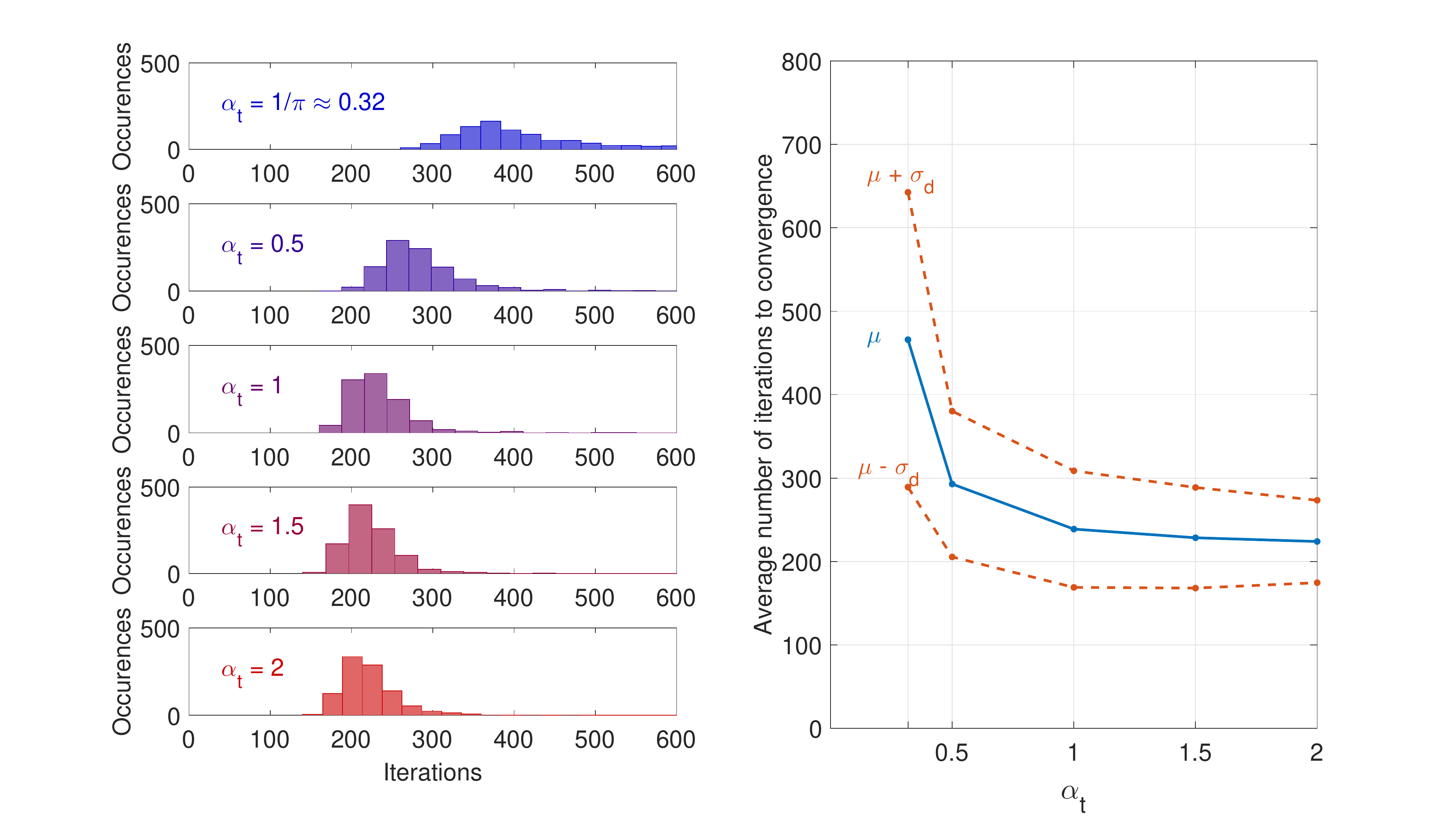}
		\caption{Statistical study of the convergence of the algorithm according to the factor $\alpha_t$. The results are gathered in the right-hand side graphic, presenting the average $\mu$ of each distribution and the standard deviation $\sigma_d$. 1000 trials are computed for each value of $\alpha_t$.}
		\label{fig:ConvAlphat}
	\end{figure}

	Finally, a convergence study is presented considering the impact of $\alpha_t$ acting on the frequency oversampling. A set of numerical simulations are computed with SNR = $10^6$ and sampling $m = 6n$, considering 1000 trials for each value of $\alpha_t$~(Fig.~\ref{fig:ConvAlphat}).
	
	This numerical study depicts the relation between the number of iterations required to reach convergence with respect to $\alpha_t$, demonstrating the positive impact of a high quality factor and a fine frequency sampling on the convergence speed.
	
	\section{Pratical implementation and experimental results}
	
\begin{figure}[h!]
		\centering
		\includegraphics[width=0.65\textwidth]{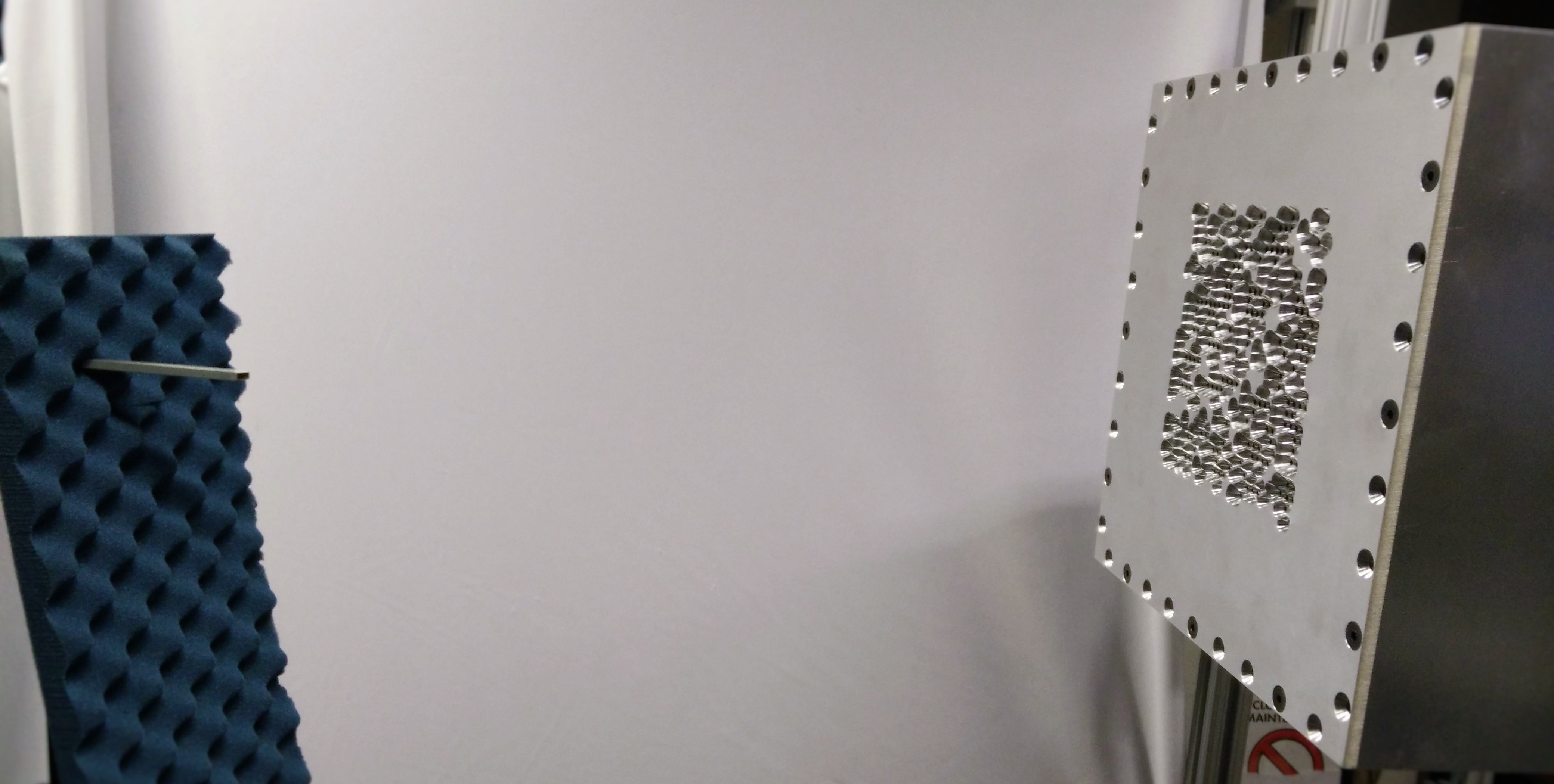}
		\caption{Radiating metasurface implemented for the validation of the proposed phaseless computational technique.}
		\label{fig:experimentalsetup}
	\end{figure}

	The theory of phaseless computational imaging is experimentally validated using a metasurface operating in the microwave regime. To this end, a metallic leaky cavity of $28.5 \times 28.5 \times 15.2\ \text{cm}^3$ was created (Fig.~\ref{fig:experimentalsetup}), inspired by a computational imaging prototype introduced in~\cite{fromenteze2015computational}.
	
\begin{figure}[h!]
		\centering
		\includegraphics[width=0.42\textwidth]{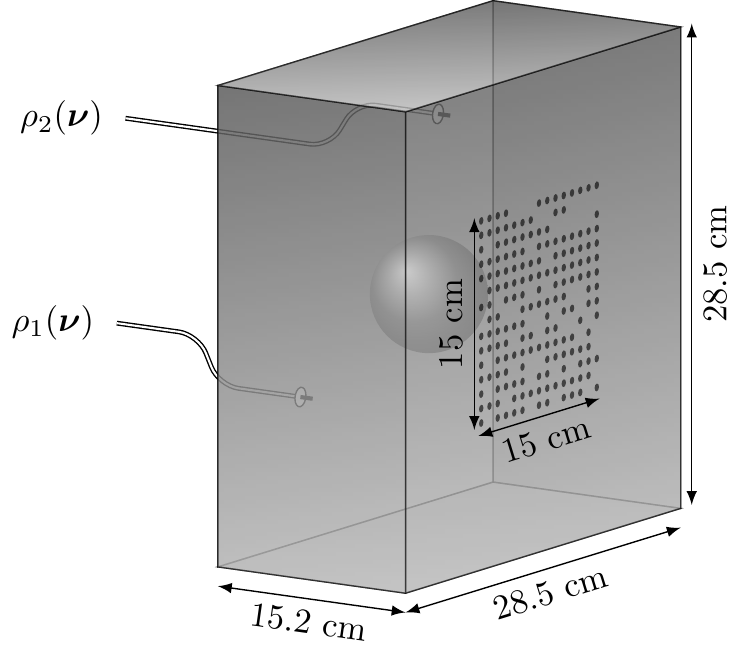}
		\caption{Radiating metasurface implemented for the validation of the proposed phaseless computational technique.}
		\label{fig:CavityExp}
	\end{figure}

	The front plate is perforated by a  $15 \times 15\ \text{cm}^2$ square array of circular holes randomly set on a $0.6\ \text{cm}$ uniform grid. An open-ended waveguide source is set in front of the cavity and localized using the computational system. In contrast with the setup depicted in Fig.~\ref{fig:Cavity}, two ports are connected to the back of the cavity~(Fig.~\ref{fig:CavityExp}). In this way, different superpositions of modes can be measured by the two ports according to their locations in the cavity, increasing the amount of independent information in the frequency domain. A spherical reflector is also set in the cavity, providing additional mode mixing and increasing the number of uncorrelated states in the cavity~\cite{fromenteze2015computational,montaldo2005building}.

\begin{figure}[h!]
		\centering
		\includegraphics[width=0.93\textwidth]{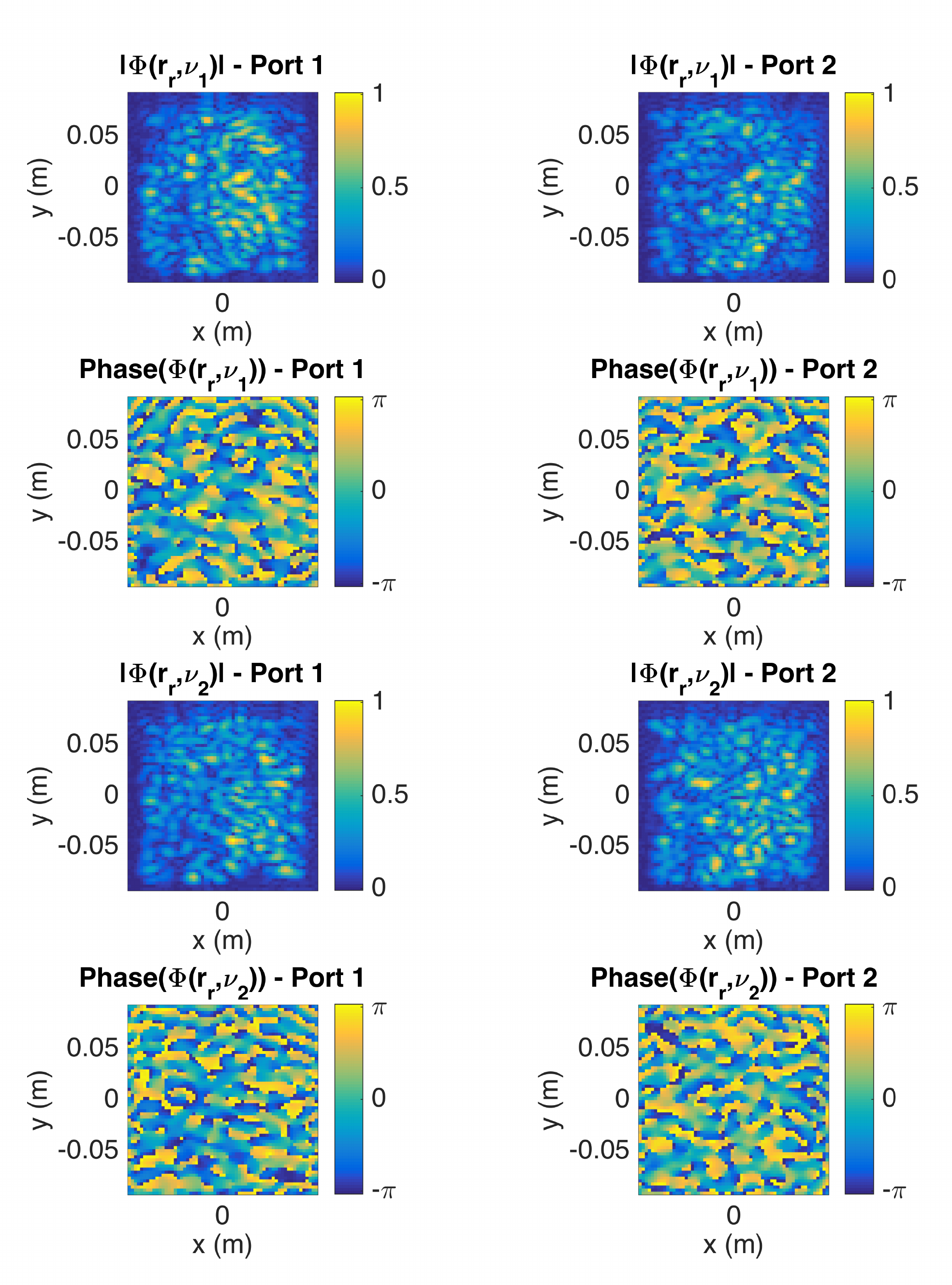}
		\caption{Comparison of the near-field distributions $\Phi_1(r_r,\nu)$ and $\Phi_2(r_r,\nu)$ measured for  the independent excitation of ports 1 and 2. The results are depicted for two consecutive frequency $\nu_1 =  23$~GHz and $\nu_2 = 23.002$ GHz of the frequency vector $\bm \nu$.}
		\label{fig:Patterns}
	\end{figure}

	To match the simulations, the operating frequency range is defined in the K-band between $\nu_{min} = 17.5$~GHz and $\nu_{max} = 26.5$~GHz, sampled by $m_\nu = 3601$ frequency points. In this way, $m = 2 m_\nu$ points are measured for the phaseless estimation of the $n$ voxels constituting $\bm f$. The patterns radiated at each frequency and for each excitation port are measured by a near-field, single-polarized probe moved on a planar synthetic aperture by a translation stage. This field is numerically back-propagated to the metasurface plane to estimate $\Phi_1(\bm r_r, \bm \nu)$ and $\Phi_2(\bm r_r, \bm \nu)$, the transfer functions of the cavity when exciting ports 1 and 2, respectively. Examples of near-field distributions are presented in Fig.~(\ref{fig:Patterns}) for two consecutive frequencies of the vector $\bm \nu$. Different pseudo-random spatial fields distributions are thus obtained as a function of the excited port and of the frequency, due to the low level of loss of this cavity.
	
	The measured quality factor of the cavity is about $12 000$ for both ports, determined by fitting exponential functions to the measured radiation patterns expressed in the time domain using a Fourier transform and taking the root mean square over the spatial dimension $\bm r_r$. A formulation matching Eq.~\ref{eq:init} is proposed according to the measured signals $\bm \rho_1 \in \mathbb{C}^{m_\nu}$ and $\bm \rho_2  \in \mathbb{C}^{m_\nu}$ on ports 1 and 2 by concatenating the measured signals and the corresponding sensing matrices computed with Eq.~\ref{eq:Hmat}:
	
	\begin{align}
	&\bm \rho = [\bm \rho_1, \bm \rho_2]\\		
	&\bm H    = [ \bm H_{1} , \bm H_2]
	\end{align}
	
	where $\bm \rho \in \mathbb{C}^{m}$ and $\bm H \in \mathbb{C}^{m \times n}$, dimensioned such that the target field distribution $\bm f \in \mathbb{C}^n$ is estimated, with $m = 2 m_\nu$. The upper bound of the number of reconstructed voxels can be computed according to the quality factor and the setup presented in this experiment. Merging Eqs.~\ref{eq:tau} and~\ref{eq:Q} gives:
	
	\begin{equation}
	\tau = \frac{Q}{\pi \nu_0}
	\end{equation}
	
	From the expression of the time constant $\tau$, the maximum number of orthogonal channels is bounded by~\cite{marks2016spatially}:
	
	\begin{align}
	m_\nu &< \pi \tau B\\
	&< Q \frac{B}{\nu_0} \approx 4900
	\end{align}
	
	In the considered experiment, there are at most $m = 2 m_\nu = 9800$ independent frequency points, measured on the two ports of the cavity. Under the assumption of an ideal estimation of the sensing matrix $\bm H$ and a sampling $n = m/5$, a maximum of $1960$ voxels can be reconstructed. For this validation, since $m_\nu =  3601$ frequency points are measured, a maximum of $n = 1440$ voxels can be estimated with a sampling of $m = 5n$. Considering this setup, the value of the parameter $\alpha_t$ is given for one port by Eq.~\ref{eq:Q}:
	
	\begin{align}
	\alpha_t = \frac{Q d\nu}{\pi \nu_0} = \frac{Q B}{\pi \nu_0 m_\nu} \approx 4.3 
	\end{align}
	
	Under these conditions and considering that two signals are measured for the estimation of $\hat{\bm f_I}$, a fast convergence of the iterative phase retrieval algorithm is expected. 
	
\begin{figure}[h!]
		\centering
		\includegraphics[width=\widthfig\textwidth]{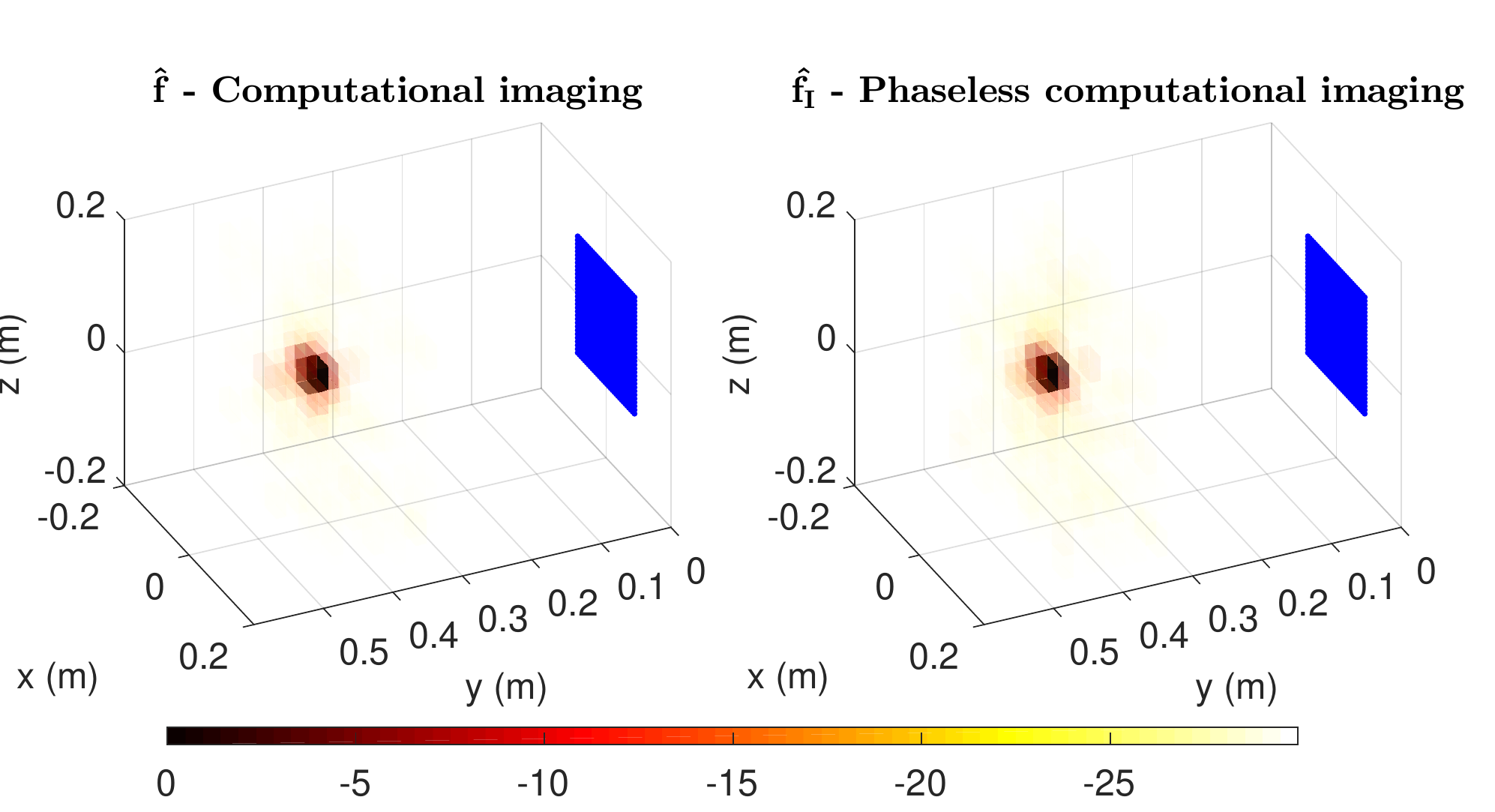}
		\caption{Localization of a field source on a domain of $10 \times 10 \times 10$~voxels, with and without the phase information. The blue square represents the array of equivalent dipoles constituting the radiating metasurface.}
		\label{fig:C4_3D_small}
	\end{figure}

	A first validation is proposed for a scene made of $10 \times 10 \times 10 = 1000$ voxels centered around \mbox{$x=0, y = 0.4 \text{ m}, z = 0$}. In accordance with the numerical study, the spatial sampling is defined as $d_x = \lambda_c R/D_x \approx 3.6$ cm, $d_z = \lambda_c R/D_z \approx 3.6$~cm and $d_y = c/(2B) \approx 1.7$~cm, with $R = 0.4$~m, \mbox{$D_x = D_z = 0.15$~cm}, and $B = 9$~GHz. A probe is first set in front of the radiating metasurface in the middle of the pre-defined region of interest. A comparison between the retrieved field distributions $\hat{\bm f}$~and $\hat{\bm f_I}$~is presented in Fig.~\ref{fig:C4_3D_small}.
	
\begin{figure}[h!]
		\centering
		\includegraphics[width=\widthfig\textwidth]{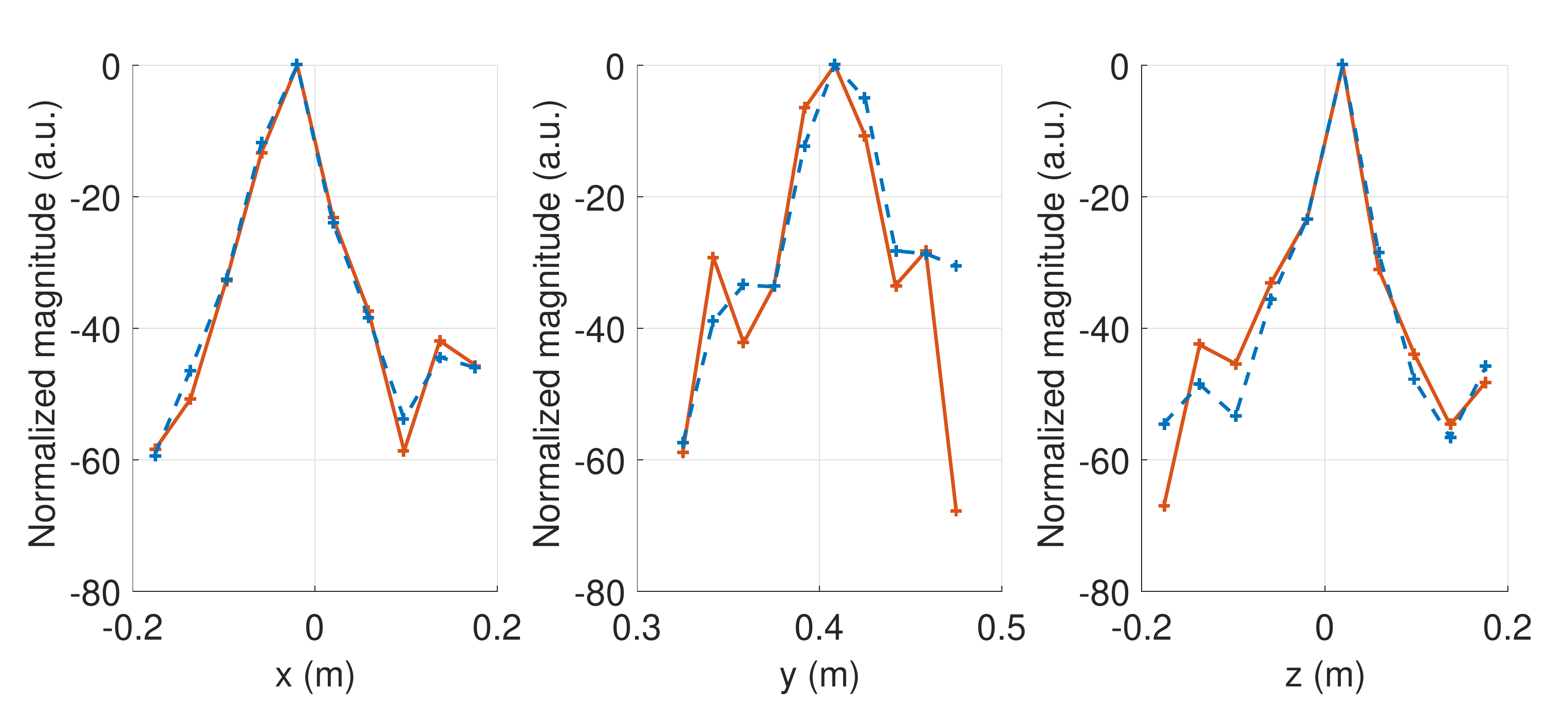}
		\caption{Comparison of the $x$, $y$, and $z$-cuts extracted at the maximum value of the reconstructed fields $\hat{\bm f}$~and $\hat{\bm f_I}$. The orange solid lines correspond to the phaseless results $\hat{\bm f}_I$, and are compared to the dashed blue lines standing for the reconstructions from complex measurements $\hat{\bm f}$.}
		\label{fig:C4_small}
	\end{figure}

	The field distribution reconstructed from intensity-only measurements matches the estimation from the complex measurements, taken as a reference in this study. A higher level of noise is observed in the phaseless case, most likely due to a non-ideal estimation of the sensing matrix $\bm H$ considering that a supplementary mounting structure (source of diffraction) was added after the near-field characterization of the leaky cavity leading to a relative error $\epsilon = 0.57$. A more precise comparison of the two estimated fields $\hat{\bm f}$ and $\hat{\bm f_I}$ is proposed, extracting the $x$, $y$, and $z$-cuts from the maximum values (Fig.~\ref{fig:C4_small}).
	
\begin{figure}[H]
		\centering
		\includegraphics[width=\widthfig\textwidth]{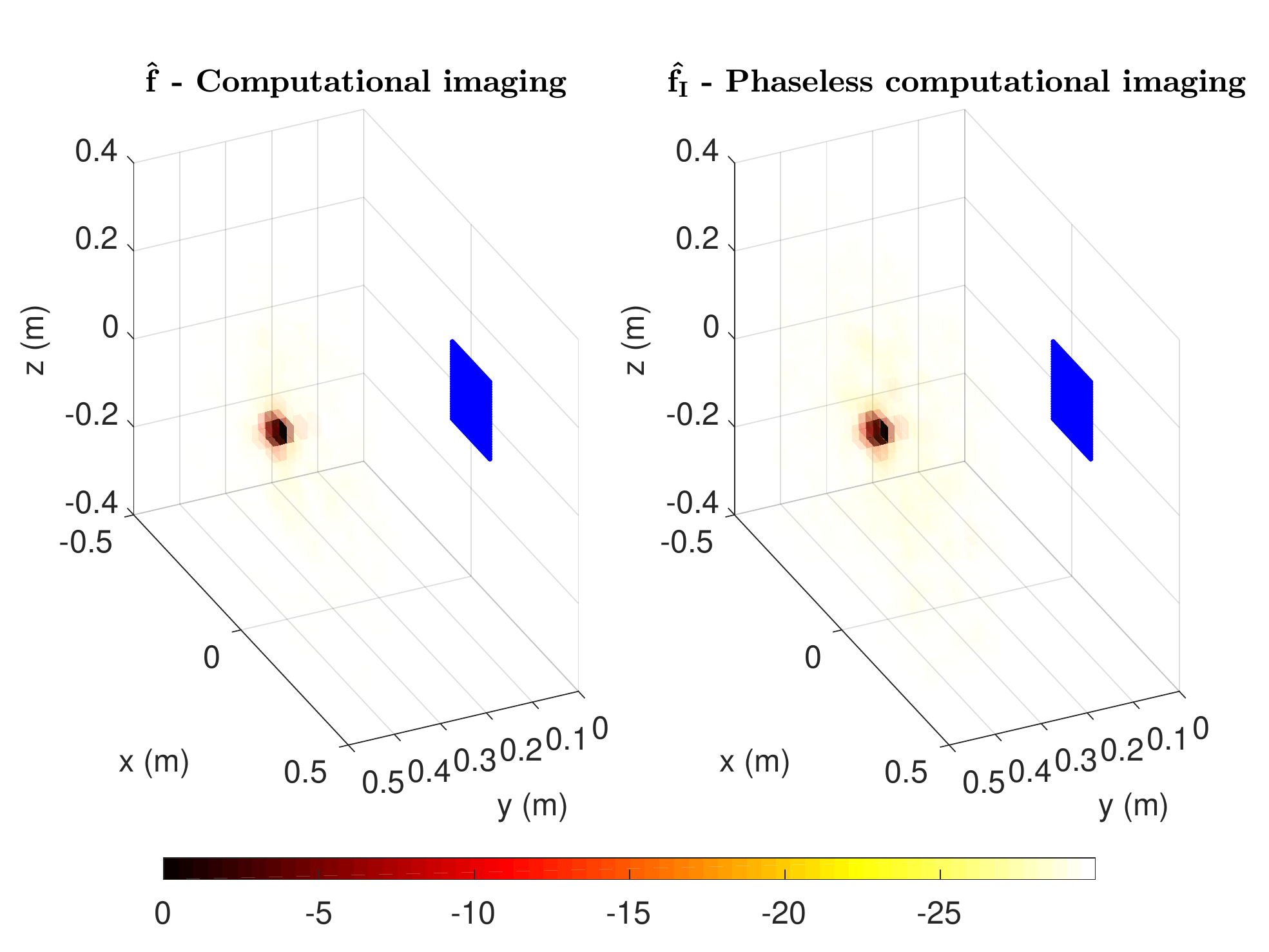}
		\caption{Localization of a field source set in front of the radiating metasurface in a domain of $20 \times 20 \times 10$ voxels, with and without the phase information. The blue square represents the array of equivalent dipoles constituting the radiating metasurface.}
		\label{fig:C4_3D}
	\end{figure}

	Comparable locations and resolutions are observed in both cases, validating the fidelity of the truncated Wirtinger flow applied to this phaseless computational system. A larger domain is considered for the last part of this experimental validation, studying the impact of the sampling $m/n$ presented earlier in a practical situation. A domain of $20 \times 20 \times 10 = 4000$ voxels is thus considered this time, conserving the same spatial sampling and centered at the same location. We now consider a sampling of $m/n = 9800/4000 = 2.45$ (with the approximation of two independent measurements $\bm \rho_1$ and $\bm \rho_2$). In contrast with the numerical simulations where spatial random field distributions were considered, the experimental cases are focused on the reconstruction of a punctual point like object. Even if a relative error of $\epsilon < 10^{-5}$ may not be reachable, we are interested in determining whether a localization of the field source is possible in the given conditions. A comparison of the reconstructions achieved with and without the phase information is once again presented from the same measurements as in the previous case (Fig.~\ref{fig:C4_3D}).
	
\begin{figure}[h!]
		\centering
		\includegraphics[width=\widthfig\textwidth]{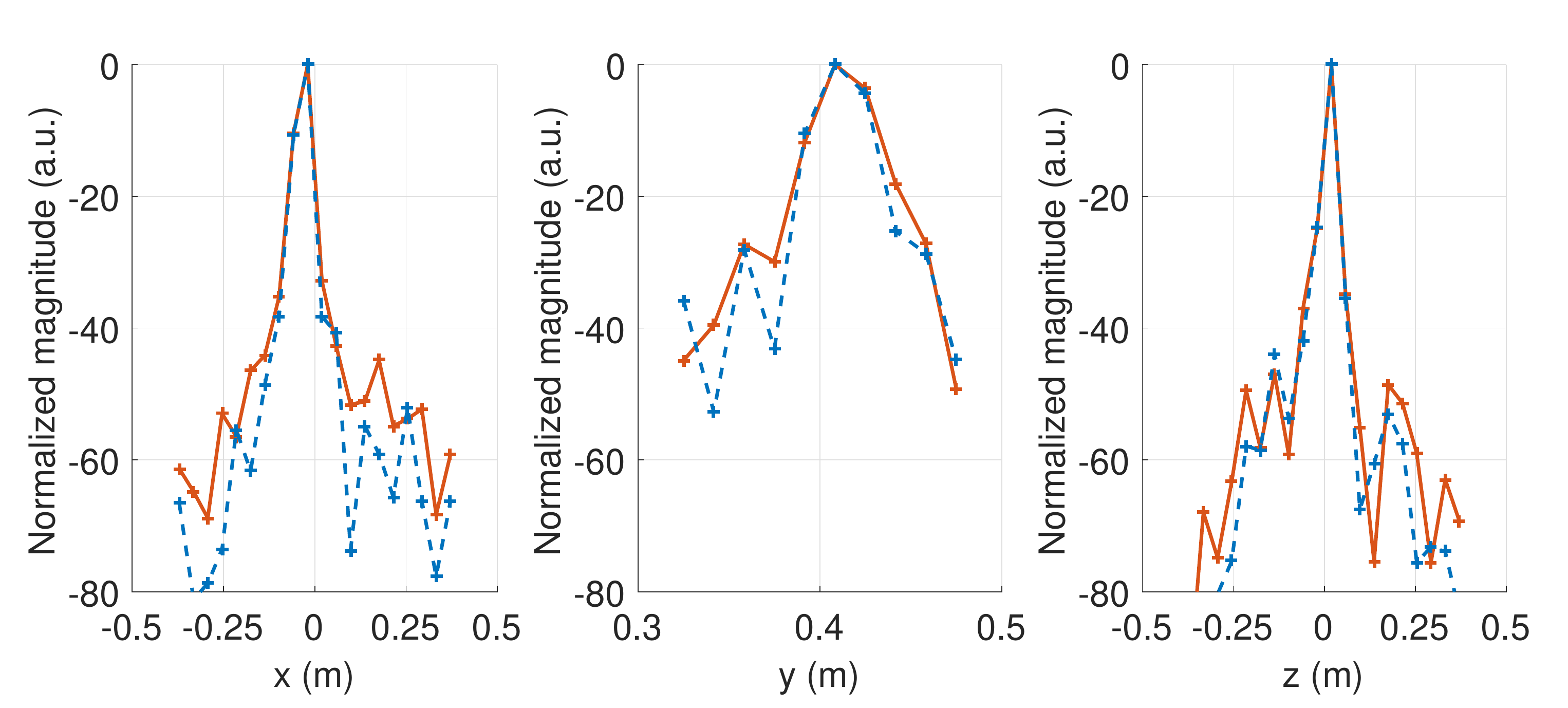}
		\caption{Comparison of the $x$, $y$, and $z$-cuts extracted at the maximum value of the reconstructed fields $\hat{\bm f}$ and $\hat{\bm f_I}$. The orange solid lines correspond to the phaseless results $\hat{\bm f}_I$, and are compared to the dashed blue lines standing for the reconstructions from complex measurements $\hat{\bm  f}$.}
		\label{fig:C4}
	\end{figure}

 A relative error of $\epsilon = 0.82$ is obtained for this phaseless reconstruction, which is consistent with expectations of a larger $\epsilon$ than in the previous case due to the lower sampling $m/n$ in this case. Despite this larger error, the localization of the field source remains possible, as depicted in the comparison between the field cuts presented in Fig.~\ref{fig:C4}. 
	
\begin{figure}[h!]
		\centering
		\includegraphics[width=\widthfig\textwidth]{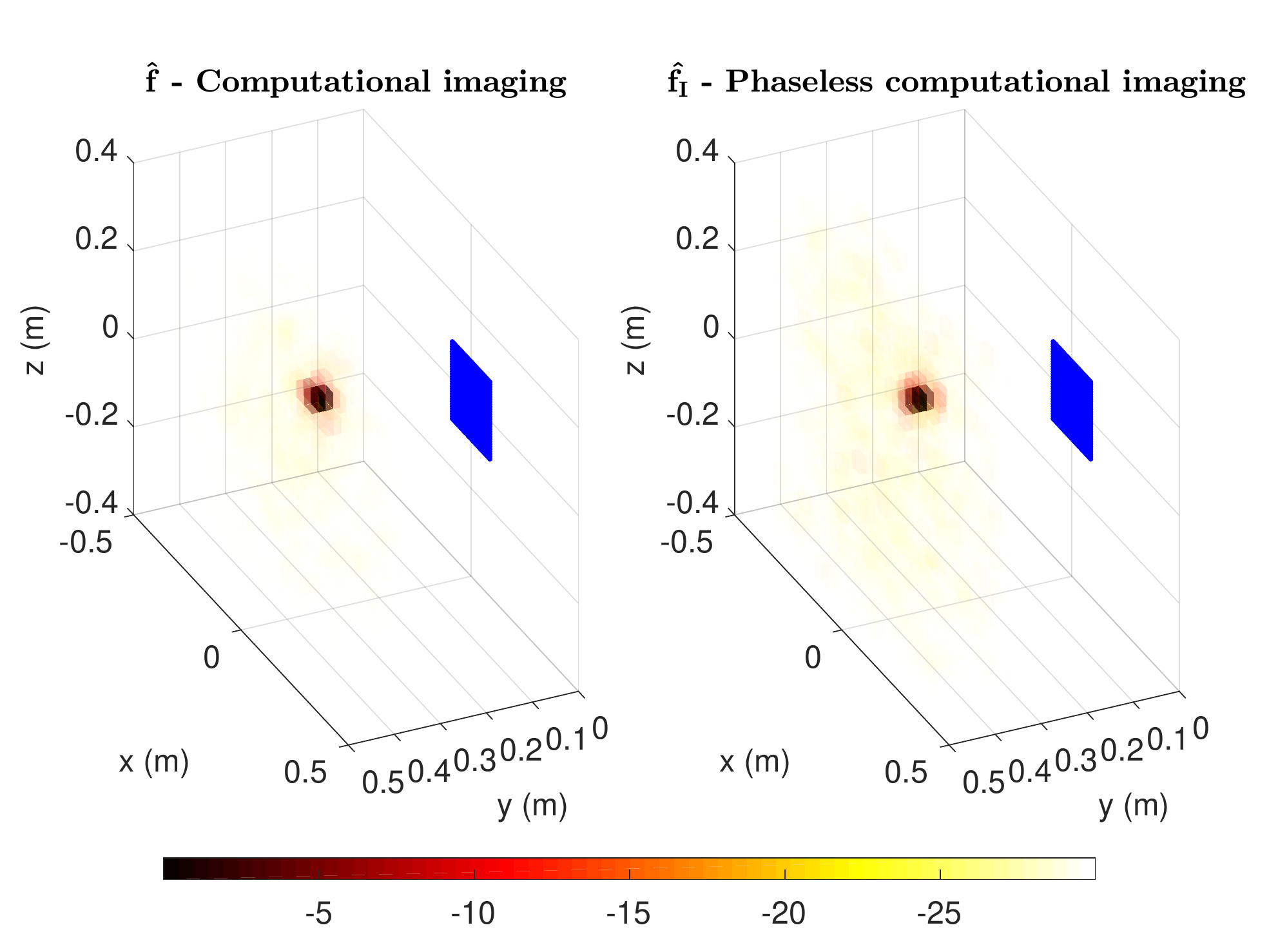}
		\caption{Localization of a field source shifted from the center in a domain of $20 \times 20 \times 10$ voxels, with and without the phase information. The blue square represents the array of equivalent dipoles constituting the radiating metasurface.}
		\label{fig:TL4_3D}
	\end{figure}

\begin{figure}[h!]
		\centering
		\includegraphics[width=\widthfig\textwidth]{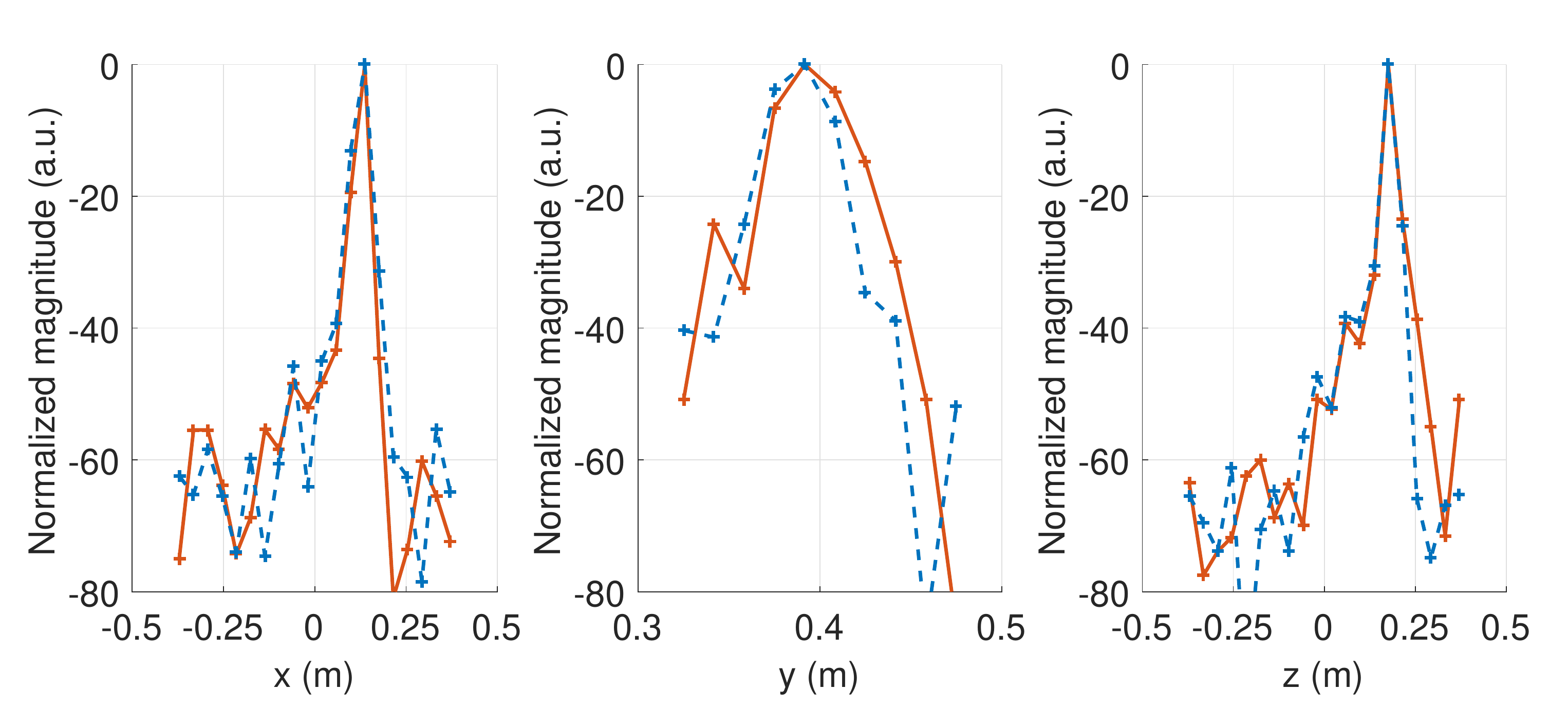}
		\caption{Comparison of the $x$, $y$, and $z$-cuts extracted at the maximum value of the reconstructed fields $\hat{\bm f}$ and $\hat{\bm f_I}$ for a source field shifted from the center of the imaging domain. The orange solid lines correspond to the phaseless results $\hat{\bm f}_I$, and are compared to the dashed blue lines standing for the reconstructions from complex measurements $\hat{\bm  f}$.}
		\label{fig:TL4}
	\end{figure}

	Taking benefit of this larger domain, the source is translated at an off-center location to be imaged (Fig.~\ref{fig:TL4_3D}). With a relative error of $\epsilon = 1.03$, the three-dimensional reconstructions with and without the phase information remain comparable. The $x$, $y$, and $z$-cuts are extracted from the maximum value of both reconstructions for a finer analysis~(Fig.~\ref{fig:TL4}).\\
	
	While the fields extracted from the transverse axis are almost identical, a discrepancy is noted in the range axis. Indeed, the range information is coded by the time of arrival of propagating waves, mainly represented by phase ramps in the frequency domain that cannot be exploited in intensity measurements. Despite the lack of phase data and the under-sampling of the considered case, it remains possible to estimate the location of the field source with a degraded resolution compared to the computational imaging case based on complex valued measurements.
	
	\section{Conclusion}
	
	An application of a phase retrieval algorithm to a computational imaging system has been presented, allowing for the spatial reconstruction of field distributions from phaseless measurements. The truncated Wirtinger flow has been adapted in this study to determine the position of field sources from the measurements of a metasurface radiating pseudo-orthogonal patterns in the frequency domain. In contrast with the coded X-ray diffraction experiments simulated by Cand\`es et al., there is no need of a reconfigurable random lens since the information is coded in the frequency domain by a static and passive device. While this application has been presented in the microwave range to where it is possible to compare the results with and without the phase information, the most useful applications stand at higher frequencies where the burdensome measurement of phase is problematic and limits the realization of 3D imaging systems. Future studies will thus focus on extending the implementation of such a technique to the terahertz, visible and infrared domains.
	
	\section*{Aknowledgements}
	This work was supported by the Air Force Office of Scientific Research (AFOSR, Grant No. FA9550-12-1-0491).
	
\end{document}